\documentclass[twocolumn,showpacs,amsmath,amssymb,letter]{revtex4}

\usepackage{graphicx}
\usepackage{dcolumn}
\usepackage{bm}

\begin{document}

\preprint{APS/123-QED}

\title{Local and Nonlocal Strain Rate Fields and Vorticity Alignment in Turbulent Flows}

\author{Peter E. Hamlington$^1$\footnote{email: peterha@umich.edu (corresponding author)},
        J\"org Schumacher$^2$\footnote{email: joerg.schumacher@tu-ilmenau.de}, and
        Werner J. A. Dahm$^1$\footnote{email: wdahm@umich.edu}}
 \affiliation{$^1$ Laboratory for Turbulence \& Combustion (LTC),
 Department of Aerospace Engineering, The University of Michigan, Ann Arbor, MI 48109-2140, USA\\
 $^2$ Department of Mechanical Engineering, Technische Universit\"at Ilmenau, D-98684 Ilmenau, Germany}

\date{\today}

\begin{abstract}
Local and nonlocal contributions to the total strain rate tensor
$S_{ij}$ at any point $\textbf{x}$ in a flow are formulated from
an expansion of the vorticity field in a local spherical
neighborhood of radius $R$ centered on $\textbf{x}$. The resulting
exact expression allows the nonlocal (background) strain rate
tensor $S_{ij}^{B}(\textbf{x})$ to be obtained from
$S_{ij}(\textbf{x})$. In turbulent flows, where the vorticity
naturally concentrates into relatively compact structures, this
allows the local alignment of vorticity with the most extensional
principal axis of the background strain rate tensor to be
evaluated.  In the vicinity of any vortical structure, the
required radius $R$ and corresponding order $n$ to which the
expansion must be carried are determined by the viscous
lengthscale $\lambda_{\nu}$.  We demonstrate the convergence to
the background strain rate field with increasing $R$ and $n$ for
an equilibrium Burgers vortex, and show that this resolves the
anomalous alignment of vorticity with the intermediate eigenvector
of the total strain rate tensor.  We then evaluate the background
strain field $S_{ij}^{B}(\textbf{x})$ in DNS of homogeneous
isotropic turbulence where, even for the limited $R$ and $n$
corresponding to the truncated series expansion, the results show
an increase in the expected equilibrium alignment of vorticity
with the most extensional principal axis of the background strain
rate tensor.
\end{abstract}

\pacs{47.27.-i,47.32.C-,47.27.De}
\maketitle

\section{\label{sec:level1}Introduction}

Vortex stretching is the basic mechanism by which kinetic energy
is transfered from larger to smaller scales in three-dimensional
turbulent flows
\cite{Burgers1948,Batchelor1964,Lundgren1982,Lundgren1993}. An
understanding of how vortical structures are stretched by the
strain rate field $S_{ij}({\bf x})$ is thus essential to any
description of the energetics of such flows. Over the last two
decades, direct numerical simulations (DNS)
\cite{Ashurst1987,She1991,Nomura1998} and experimental studies
\cite{Tsinober1992,Buch1996,Su1996,Zeff2003,Mullin2006} of the
fine-scale structure of turbulence have revealed a preferred
alignment of the vorticity with the intermediate eigenvector of
the strain rate tensor. This result has been widely regarded as
surprising. Indeed the individual components of the inviscid
vorticity transport equation, in a Lagrangian frame that remains
aligned with the eigenvectors of the strain rate tensor, are
simply
\begin{equation}\label{bs0}
  \frac{\mbox{D}\omega_1}{\mbox{D}t} = s_1 \omega_1\,,\;\;\;
  \frac{\mbox{D}\omega_2}{\mbox{D}t} = s_2 \omega_2\,,\;\;\;
  \frac{\mbox{D}\omega_3}{\mbox{D}t} = s_3 \omega_3\,,
  \end{equation}
where $s_1$, $s_2$ and $s_3$ are the eigenvalues of $S_{ij}$. For
incompressible flow, $s_1+s_2+s_3\equiv 0$, and then denoting
$s_1\ge s_2\ge s_3$ requires $s_1\geq0$ and $s_3\leq 0$. As a
consequence, (\ref{bs0}) would predict alignment of the vorticity
with the eigenvector corresponding to the most extensional
principal strain rate $s_1$. Yet DNS and experimental studies have
clearly shown that the vorticity instead is aligned with the
eigenvector corresponding to the {\it intermediate} principal
strain rate $s_2$.

A key to understanding this result is that, owing to the
competition between strain and diffusion, the vorticity in
turbulent flows naturally forms into concentrated vortical
structures. It has been noted, for example in Refs.
\cite{Nomura1998,Jimenez1992,Brasseur2005}, that the anomalous
alignment of the vorticity with the strain rate tensor
$S_{ij}({\bf x})$ might be explained by separating the local
self-induced strain rate field created by the vortical structures
themselves from the background strain field in which these
structures reside. The total strain rate tensor is thus split into
\begin{equation}\label{bs1}
S_{ij}(\textbf{x}) = S^R_{ij}(\textbf{x})+S^B_{ij}(\textbf{x}) \:,
\end{equation}
where $S_{ij}^R$ is the {\it local} strain rate induced by a
vortical structure in its neighboring vicinity, and $S_{ij}^B$ is
the {\it nonlocal} background strain rate induced in the vicinity
of the structure by all the remaining vorticity. The vortical
structure would then be expected to align with the principal axis
corresponding to the most extensional eigenvalue of the background
strain rate tensor $S_{ij}^B({\bf x})$.
\begin{figure}[t]
\includegraphics[width=3.3in]{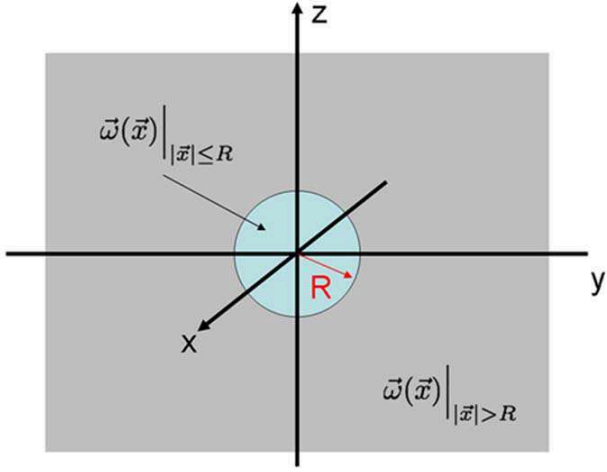}
\caption{Decomposition of the vorticity field in the vicinity of any point $\textbf{x}$ into
local and nonlocal parts; the Biot-Savart integral in (\ref{bs6}) over each part gives
the local and nonlocal (background) contributions to the total strain rate tensor
$S_{ij}$ at $\textbf{x}$.}
\label{plane}
\end{figure}

In the following, we extend this idea and suggest a systematic
expansion of the total strain rate field $S_{ij}({\bf x})$ that
allows the background strain rate field $S_{ij}^B({\bf x})$ to be
extracted. Our approach is based on an expansion of the vorticity
over a local spherical region of radius $R$ centered at any point
${\bf x}$. This leads to an exact operator that provides direct
access to the background strain rate field. The operator is tested
for the case of a Burgers vortex, where it is shown that the local
self-induced strain field produced by the vortex can be
successfully removed, and the underlying background strain field
can be increasingly recovered as higher order terms are retained
in the expansion. The anomalous alignment of the vorticity with
respect to the eigenvectors of the total strain field is shown in
that case to follow from a local switching of the principal strain
axes when the vortex becomes sufficiently strong relative to the
background strain. Finally, the operator is applied to obtain
initial insights into the background strain $S_{ij}^B({\bf x})$ in
DNS of homogeneous isotropic turbulence, and used to compare the
vorticity alignment with the eigenvectors of the total strain
field and of this background strain field.

\section{The Background Strain Field}
The velocity $\textbf{u}$ at any point $\textbf{x}$ induced by the vorticity
field $\bm{\omega}(\textbf{x})$ is given by the Biot-Savart
integral
\begin{equation}\label{bs1a}
  \textbf{u}(\textbf{x}) = \frac{1}{4\pi} \int_{\bm{\Lambda}}
  \bm{\omega}(\textbf{x}^{\prime})\times\frac{\textbf{x}-\textbf{x}^{\prime}}
  {|\textbf{x}-\textbf{x}^{\prime}|^3} d^3 \textbf{x}^{\prime}\:,
  \end{equation}
  where the integration domain $\bm{\Lambda}$ is taken to be
  infinite or periodic.
In index notation (\ref{bs1a}) becomes
\begin{equation}\label{bs2}
  u_i(\textbf{x})= \frac{1}{4\pi} \int_{\bm{\Lambda}}
  \epsilon_{ilk} \omega_l(\textbf{x}^{\prime})\frac{\left(x_k-x_k^{\prime}\right)}
  {|\textbf{x}-\textbf{x}^{\prime}|^3} d^3 \textbf{x}^{\prime}\:,
\end{equation}
where $\epsilon_{ilk}$ is the cyclic permutation tensor. The
derivative with respect to $x_j$ gives the velocity gradient
tensor
\begin{equation}\label{bs3}
  \frac{\partial }{\partial x_j}u_i(\textbf{x}) = \frac{1}{4\pi}
  \int_{\bm{\Lambda}} \epsilon_{ilk} \omega_l
  (\textbf{x}^{\prime})\left[
  \frac{\delta_{kj}}{r^3} - 3\frac{r_k r_j}
  {r^5}\right]d^3 \textbf{x}^{\prime}\:,
\end{equation}
where $r\equiv|\textbf{x}-\textbf{x}^{\prime}|$ and $r_m \equiv
x_m-x_m^{\prime}$. The strain rate tensor $S_{ij}$ at $\textbf{x}$
is the symmetric part of the velocity gradient, namely
\begin{equation}\label{bs4}
  S_{ij}(\textbf{x})\equiv \frac{1}{2}\left(
  \frac{\partial u_i}{\partial x_j}+
  \frac{\partial u_j}{\partial x_i}\right)\:.
  \end{equation}
  From (\ref{bs3}) and (\ref{bs4}), $S_{ij}(\textbf{x})$ can be expressed \cite{Ohkitani1994} as an integral over the vorticity field as
  \begin{equation}\label{bs5}
  S_{ij}(\textbf{x})=\frac{3}{8\pi}
  \int_{\bm{\Lambda}}
  \left(\epsilon_{ikl}r_j+r_i\epsilon_{jkl}\right)\frac{r_k}{r^5}
   \omega_l(\textbf{x}^{\prime})d^3\textbf{x}^{\prime}\:.
\end{equation}
As shown in Fig.\ \ref{plane}, the total strain
rate in (\ref{bs5}) is separated into the local contribution induced by the vorticity
within a spherical region of radius $R$ centered on the point $\textbf{x}$ and the remaining
nonlocal (background) contribution induced by all the vorticity
outside this spherical region.
The strain rate tensor in (\ref{bs5}) thus becomes
\begin{eqnarray}\label{bs6}
  S_{ij}(\textbf{x}) = & \underbrace{\frac{3}{8\pi}\int_{r\leq R}\left[\cdots\right]
  d^3\textbf{x}^{\prime}} +& \underbrace{\frac{3}{8\pi}\int_{r> R} \left[\cdots
  \right] d^3\textbf{x}^{\prime}}\:, \\
  & \equiv S_{ij}^R(\textbf{x}) & \qquad \equiv S_{ij}^B(\textbf{x}) \nonumber
\end{eqnarray}
where $[\cdots]$ denotes the integrand in (\ref{bs5}).
The nonlocal background strain tensor at $\textbf{x}$ is then
\begin{equation}\label{bs7}
  S^B_{ij}(\textbf{x}) = S_{ij}(\textbf{x}) - S_{ij}^R
  (\textbf{x})\:.
\end{equation}
The total strain tensor $S_{ij}(\textbf{x})$ in (\ref{bs7}) is readily evaluated via (\ref{bs4})
from derivatives of the velocity field at point $\textbf{x}$.
Thus all that is required to obtain the background (nonlocal) strain rate tensor
$S_{ij}^B(\textbf{x})$ via (\ref{bs7}) is an evaluation of the local strain
integral $S_{ij}^R(\textbf{x})$ in (\ref{bs6}) produced by the
vorticity field $\omega_l(\textbf{x}^{\prime})$ within $r \leq R$ in Fig.\ \ref{plane}.

\subsection{Evaluating the Background Strain Rate Tensor}
The vorticity field within the sphere of radius $R$ can be represented by its
Taylor expansion about the center point $\textbf{x}$ as
\begin{eqnarray}\label{bs8}
  \left.\omega_l(\textbf{x}^{\prime})\right|_{r\leq R} &=&\omega_l(\textbf{x})+ \left(
  x^{\prime}_m - x_m\right) \left.\frac{\partial \omega_l}{\partial x_m}\right|_\textbf{x}
  \\ &+& \frac{1}{2}\left( x^{\prime}_m - x_m\right)\left( x^{\prime}_n - x_n\right)
  \left.\frac{\partial^2 \omega_l}{\partial x_m \partial
  x_n}\right|_\textbf{x}+\cdots\:.\nonumber
  \end{eqnarray}
Recalling that $x_m-x_m^{\prime} \equiv r_m$ and using $a_l, b_{lm}, c_{lmn},\ldots$
to abbreviate the vorticity and its derivatives at $\textbf{x}$, we can write (\ref{bs8}) as
  \begin{equation}\label{bs9}
  \left.\omega_l(\textbf{x}^{\prime})\right|_{r\leq R} \equiv a_l - r_m b_{lm} + \frac{1}{2} r_m
  r_n c_{lmn} -\cdots\:.
\end{equation}
Substituting (\ref{bs9}) in the $S_{ij}^{R}$ integral in (\ref{bs6}) and
changing the integration variable to $\textbf{r} = \textbf{x} -
\textbf{x}^{\prime}$, the strain tensor at $\textbf{x}$ produced by the
vorticity in $R$ is
\begin{eqnarray}\label{bs10}
  S^R_{ij}(\textbf{x}) &=& \frac{3}{8\pi}
  \int_{r\leq R}
  \left(\epsilon_{ikl}r_j+r_i\epsilon_{jkl}\right)\frac{r_k}{r^5}\\
   &\times&\left[ a_l - r_m b_{lm} + \frac{r_m
  r_n}{2} c_{lmn} - \cdots\right]d^3\textbf{r}\:.\nonumber
\end{eqnarray}
This integral can be solved in spherical coordinates centered on
$\textbf{x}$, with $r_1 = r\sin{\theta} \cos{\phi}$, $r_2 =
r\sin{\theta} \sin{\phi}$, and $r_3 = r\cos{\theta}$ for $r\in
[0,R]$, $\theta \in[0,\pi]$, and $\phi \in[0,2\pi)$.
To integrate (\ref{bs10}) note that
\begin{subequations}\label{bs11}
\begin{eqnarray}
  &&\int_{r\leq R} \frac{r_k r_j}{r^5} d^3 \textbf{r}
  = \frac{4\pi}{3} \delta_{jk}\int_0^R \frac{1}{r} dr\:,\\
  &&\int_{r\leq R} \frac{r_k r_j r_m}{r^5} d^3 \textbf{r}=
  0\:,\\
  &&\int_{r\leq R} \frac{r_k r_j r_m r_n}{r^5} d^3
  \textbf{r}=
  \frac{2\pi}{15} R^2 \left(\delta_{mn}\delta_{jk} + \right. \\
  &&\qquad\qquad\qquad\qquad\qquad\qquad\:\:\left.\delta_{mj}\delta_{kn}
  +\delta_{mk}\delta_{jn}\right)\:.\nonumber
\end{eqnarray}
\end{subequations}
The resulting local strain rate tensor at $\textbf{x}$ is then
\begin{eqnarray}\label{bs12}
  &&S_{ij}^R(\textbf{x}) =\frac{R^2}{40}c_{lmn}\left(\epsilon_{ijl}\delta_{mn}+\epsilon_{jil}\delta_{mn}
  +\epsilon_{inl}\delta_{mj}\right. \\
  &&\qquad\qquad\qquad\left.+ \epsilon_{jnl} \delta_{mi}
  +\epsilon_{iml}\delta_{nj} +
  \epsilon_{jml}\delta_{ni}\right)+O(R^4)\:,\nonumber
\end{eqnarray}
where the contribution from the $a_l$ term in (\ref{bs10}) is zero since $\epsilon_{ijl}=-\epsilon_{jil}$.
For the same reason the first two terms in (\ref{bs12}) also cancel, giving
\begin{eqnarray}\label{bs13}
  &&S_{ij}^R(\textbf{x}) =
  \frac{R^2}{40}c_{lmn}\left(\epsilon_{inl}\delta_{mj} + \epsilon_{jnl}
  \delta_{mi}\right. \\
  &&\qquad\qquad\qquad\qquad\left.+\epsilon_{iml}\delta_{nj} +
  \epsilon_{jml}\delta_{ni}\right)+O(R^4)\:.\nonumber
\end{eqnarray}
Recalling that $c_{lmn} = c_{lnm} \equiv \partial^2 \omega_l /\partial x_m\partial x_n$,
and contracting with the $\delta$ and $\epsilon$ in (\ref{bs13}), gives
\begin{eqnarray}\label{bs14}
  &&S_{ij}^R(\textbf{x})=\frac{R^2}{20}\left[
  \frac{\partial}{\partial x_j}\left(\epsilon_{iml}\frac{\partial \omega_l}{\partial
  x_m}\right)\right. \\
  &&\qquad\qquad\qquad\qquad\left. + \frac{\partial}{\partial x_i}\left(\epsilon_{jml}\frac{\partial \omega_l}{\partial
  x_m}\right)\right] +O(R^4)\:.\nonumber
\end{eqnarray}
Note that $\epsilon_{iml}\,\partial \omega_l/\partial x_m \equiv
\left(\nabla\times \bm{\omega}\right)_i$ and
\begin{equation}\label{bs15}
  \nabla\times\bm{\omega} =
  \nabla\times\left(\nabla\times \textbf{u}\right) = \nabla\left(\nabla\cdot
  \textbf{u}\right) - \nabla^2 \textbf{u}\:,
\end{equation}
so for an incompressible flow ($\nabla\cdot \textbf{u} \equiv 0$) the local
strain rate tensor at $\textbf{x}$ becomes
\begin{eqnarray}\label{bs16}
  S_{ij}^R(\textbf{x}) =
  -\frac{R^2}{20}\nabla^2\left(\frac{\partial u_i}{\partial x_j} +
  \frac{\partial u_j}{\partial x_i}\right)
  +O(R^4)\:.
\end{eqnarray}
From (\ref{bs7}), with $S_{ij}^{R}$ from (\ref{bs16})
we obtain the background strain tensor as
\begin{equation}\label{bs17}
  S_{ij}^B(\textbf{x}) =S_{ij}(\textbf{x})+\frac{R^2}{10}\nabla^2S_{ij}(\textbf{x})
  +O(R^4)\:.
\end{equation}
The remaining terms in (\ref{bs17}) result from the higher-order
terms in (\ref{bs9}).  The contributions from each of these can be
evaluated in an analogous manner, giving
\begin{eqnarray}\label{bs18}
  &&S_{ij}^B(\textbf{x})=\left[1+\frac{R^2}{10}\nabla^2
  +\frac{R^4}{280}\nabla^2\nabla^2+\cdots\right.\\
  &&\qquad\qquad+\left.
  \frac{3R^{2n-2}}{\left(2n-2\right)!
  (4n^2-1)}\left(\nabla^2\right)^{n-1}+\cdots\right]
  S_{ij}(\textbf{x})\:,\nonumber
\end{eqnarray}
where the terms shown in (\ref{bs18}) correspond to $n=1,2,\dots$.
The final result in (\ref{bs18}) is an operator that extracts the
nonlocal background strain rate tensor $S_{ij}^B$ at any point
$\textbf{x}$ from the total strain rate tensor $S_{ij}$. For the
Taylor expansion in (\ref{bs8}), this operator involves Laplacians
of the total strain rate field $S_{ij}(\textbf{x})$.

\subsection{Practical Implementation}
When using (\ref{bs18}) to examine the local alignment of any
concentrated vortical structure with the principal axes of the
background strain rate field $S_{ij}^B(\textbf{x})$ in which it
resides, the radius $R$ must be taken sufficiently large that the
spherical region $| {\bf x^{\prime}} - {\bf x} | \leq R$ encloses
essentially all the vorticity associated with the structure, so
that its local induced strain rate field is fully accounted for.
Generally, as $R$ increases it is necessary in (\ref{bs18}) to
retain terms of increasingly higher order $n$ to maintain a
sufficient representation of $\bm{\omega}(\textbf{x}^{\prime})$
over the spherical region. Thus for any vortical structure having
a characteristic gradient lengthscale $\lambda_{\nu}$, it can be
expected that $R$ must be of the order of $\lambda_{\nu}$, and $n$
will then need to be sufficiently large to adequately represent
the vorticity field within this sphere.  However, since the local
gradient lengthscale in the vorticity field in a turbulent flow is
determined by an equilibrium between strain and diffusion, the
vorticity field over the lengthscale $\lambda_{\nu}$ will be
relatively smooth, and thus relatively low values of $n$ may
suffice to give a usable representation of
$\bm{\omega}(\textbf{x}^{\prime})$.  This is examined in the
following Section.

\begin{figure}[b]
\includegraphics[width=2.5in]{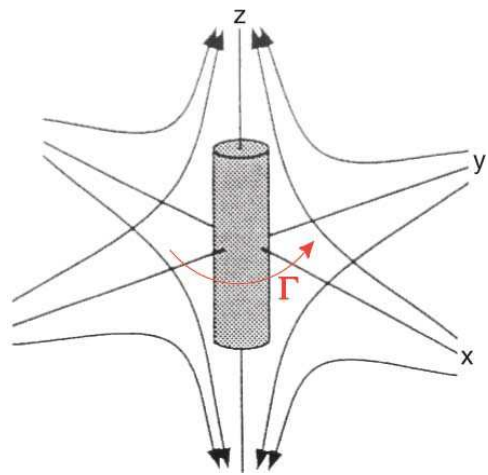}
\caption{Equilibrium Burgers vortex with circulation $\Gamma$ and
strain-limited viscous diffusion lengthscale $\lambda_{\nu}$ in a
uniform, irrotational, axisymmetric background strain rate field
$S_{ij}^{B}(\bf{x})$.} \label{burgf1}
\end{figure}

\section{Test Case: Burgers Vortex}
The equilibrium Burgers vortex
\cite{Burgers1948,Lundgren1982,Buch1996,Gibbon1999} is formed from
vorticity in a fluid with viscosity $\nu$ by a spatially uniform,
irrotational, axisymmetric background strain rate field
$S_{ij}^{B}$ that has a single extensional principal strain rate
$S_{zz}$ directed along the $z$ axis, as shown in Fig.
\ref{burgf1}.  This simple flow, often regarded as an idealized
model of the most concentrated vortical structures in turbulent
flows, provides a test case for the result in (\ref{bs18}). The
combined strain rate field $S_{ij}(\textbf{x})$ produced by the
vortex and the background strain flow should, when applied in
(\ref{bs18}), produce the underlying background strain field
$(S_{rr}^{B}, S_{\theta\theta}^{B}, S_{zz}^{B}) = (-\frac{1}{2},
-\frac{1}{2}, 1) S_{zz}$ at all $\textbf{x}$ when $R \to \infty$
and all orders $n$ are retained. For finite $R$ and $n$, the
resulting $S_{ij}^B(\textbf{x})$ will reflect the convergence
properties of (\ref{bs18}).

\subsection{Strain Rate Tensor}
The equilibrium Burgers vortex aligned with the extensional principal axis of the
background strain rate field has a vorticity field
\begin{equation}\label{burg1}
  \bm\omega(\textbf{x}) = \omega_z(r)\hat{\textbf{z}} =
  \frac{\alpha}{\pi} \frac{\Gamma}{\lambda_\nu^2}\exp\left(- \alpha \eta^2\right)\hat{\textbf{z}}\:,
\end{equation}
where $\Gamma$ is the circulation, $\lambda_\nu$ is the viscous
lengthscale that characterizes the diameter of the vortex,
$\eta\equiv r/\lambda_\nu$ is the radial similarity coordinate,
and the constant $\alpha$ reflects the chosen definition of
$\lambda_{\nu}$.  Following \cite{Buch1996}, $\lambda_{\nu}$ is
taken as the full width of the vortical structure at which
$\omega_{z}$ has decreased to one-fifth of its peak value, for
which $\alpha = 4 \ln 5$. When diffusion of the vorticity is in
equilibrium \cite{Buch1996} with the background strain, then
\begin{equation}\label{equilib1}
  \lambda_{\nu} = \sqrt{8\alpha}\left(\frac{\nu}{S_{zz}}\right)^{1/2}\:.
\end{equation}
The combined velocity field $\textbf{u}(\textbf{x})$ produced by the vortex and
the irrotational background strain is given by the cylindrical components
\begin{subequations}\label{burg2}
\begin{eqnarray}
  u_r(r,\theta,z) &=& -\frac{S_{zz}}{2} r \,,\\
  u_\theta(r,\theta,z) &=& \frac{\Gamma}{2\pi\lambda_{\nu}} \frac{1}{\eta} \left[1-\exp\left(-\alpha\eta^2\right)\right] \,,\\
  u_z(r,\theta,z) &=& S_{zz} z\,.
\end{eqnarray}
\end{subequations}

The combined strain rate tensor for such a Burgers vortex is thus
\begin{equation}\label{burg3}
  S_{ij}(\textbf{x}) = \left[\begin{array}{ccc}
                                -S_{zz}/2 & S_{r\theta}^{\smallskip v} & 0 \\
                                S_{r\theta}^{\smallskip v} & -S_{zz}/2 & 0 \\
                                0 & 0 & S_{zz}
                                \end{array}\right]\:,
\end{equation}
where $S_{r\theta}^{\smallskip v}$ is the
shear strain rate  induced by the vortex, given by
\begin{eqnarray}
  S_{r\theta}^{\smallskip v}\left(\textbf{x}\right) =
  \frac{\Gamma}{\pi\lambda_\nu^2}
  \left[\left(\alpha+\frac{1}{\eta^2}\right)
  \exp\left(-\alpha\eta^2\right) - \frac{1}{\eta^2}\right]\,.
\label{burg45}
\end{eqnarray}
From (\ref{burg3}), $S_{ij}(\textbf{x})$ has one extensional
principal strain rate equal to $S_{zz}$ along the
$\hat{\textbf{z}}$ axis, with the remaining two principal strain
axes lying in the $r$-$\theta$ plane and corresponding to the
principal strain rates
\begin{equation}\label{s23}
   s = -\frac{1}{2}S_{zz} \pm |S_{r\theta}^{\smallskip v}| \:.
\end{equation}
As long as the largest $s$ in (\ref{s23}) is smaller than
$S_{zz}$, the most extensional principal strain rate $s_{1}$ of
$S_{ij}$ will be $S_{zz}$, and the corresponding principal strain
axis will point in the $\hat{\textbf{z}}$ direction.  The
vorticity is then aligned with the most extensional eigenvector of
$S_{ij}$.  This remains the case until the vortex becomes
sufficiently strong relative to the background strain rate that $s
> S_{zz}$, namely
\begin{equation}\label{switch1}
   |S_{r\theta}^{\smallskip v}|\geq \frac{3}{2}S_{zz}\:,
\end{equation}
which from (\ref{burg45}) occurs wherever
\begin{equation}\label{axisswitch}
  \left(\alpha+\frac{1}{\eta^2}\right)\exp(-\alpha\eta^2) -
  \frac{1}{\eta^2}\geq
  \frac{3\pi}{2}\left(\frac{\Gamma/\lambda_\nu^2}{S_{zz}}\right)^{-1}\:.
\end{equation}
At any $\eta$ for which (\ref{axisswitch}) is satisfied, the most
extensional principal axis of the combined strain rate tensor
$S_{ij}(\textbf{x})$ will switch from the $\hat{\textbf{z}}$
direction to instead lie in the $r$-$\theta$ plane.  Since the
vorticity vector everywhere points in the $\hat{\textbf{z}}$
direction, wherever (\ref{axisswitch}) is satisfied the principal
axis of $S_{ij}$ that is aligned with the vorticity will switch
from the most extensional eigenvector to the intermediate
eigenvector.  This alignment switching is purely a result of the
induced strain field $S_{ij}^{\smallskip v}(\textbf{x})$ locally
dominating the background strain field $S_{ij}^{B}(\textbf{x})$.

The dimensionless vortex strength parameter
\begin{equation}\label{axisswitchparam}
   \Omega \equiv \left[\frac{\Gamma/\lambda_{\nu}^{2}}{S_{zz}}\right]
    = \frac{\pi}{\alpha} \frac{\omega_{max}}{S_{zz}}
\end{equation}
on the right-hand side of (\ref{axisswitch}) characterizes the
relative strength of the background strain and the induced strain
from the vortical structure, where $\omega_{max}$ is obtained from
(\ref{burg1}) at $\eta=0$.  For
\begin{equation}
\Omega < \Omega^{*} \approx 2.45\,,
\end{equation}
the background strain rate $S_{zz}$ is everywhere larger than the
largest $s$ in (\ref{s23}), and thus no alignment switching occurs
at any $\eta$. For $\Omega > \Omega^{*}$, alignment switching will
occur over the limited range of $\eta$ values that satisfy
(\ref{axisswitch}). With increasing values of $\Omega$, more of
the vorticity field will be aligned with the intermediate
principal axis of the $\textit{combined}$ strain rate tensor, even
though all of the vorticity field remains aligned with the most
extensional principal axis of the $\textit{background}$ strain
rate tensor.

\begin{figure}[t]
\includegraphics[width=3.5in]{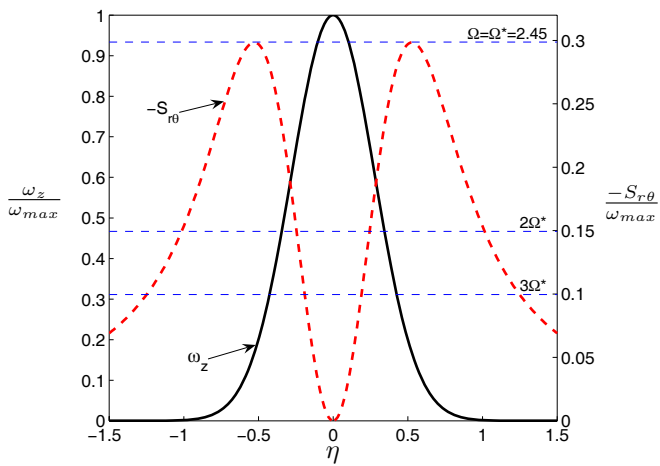}
\caption{Similarity profiles of $\omega_{z}(\eta)$ and
$S_{r\theta}(\eta)$ for any equilibrium Burgers vortex; wherever
$-S_{r\theta}$ exceeds the horizontal line determined by the
relative vortex strength parameter $\Omega$ in
(\ref{axisswitchparam}) the most extensional principal axis of the
total strain rate $S_{ij}(\bf{x})$ switches from the
$\hat{\textbf{z}}$-axis to lie in the $r$-$\theta$ plane.}
\label{burgf2}
\end{figure}

Figure \ref{burgf2} shows the vorticity $\omega_z$ and the induced
shear strain component $-S_{r\theta}^{\smallskip v}$ as a function
of $\eta$. The horizontal dashed lines correspond to three
different values of $\Omega$, and indicate the range of $\eta$
values where the alignment switching in (\ref{axisswitch}) occurs
for each $\Omega$. Wherever $-S^{\smallskip v}_{r\theta}$ is above
the dashed line for a given $\Omega$, the vorticity will be
aligned with the local intermediate principal axis of the combined
strain rate field.

In principle, regardless of the vortex strength parameter
$\Omega$, at any $\eta$ the result in (\ref{bs18}) can reveal the
alignment of the vorticity with the most extensional principal
axis of the background strain field $S_{ij}^B$. However, this
requires $R$ to be sufficiently large that a sphere with diameter
$2R$, centered at the largest $\eta$ for which $-S^{\smallskip
v}_{r\theta}$ in Fig.\ \ref{burgf2} is still above the horizontal
dashed line, will enclose essentially all of the vorticity
associated with the vortical structure.  As $\Omega$ increases,
the required $R$ will increase accordingly as dictated by
(\ref{axisswitch}), and as $R$ is increased the required $n$ in
(\ref{bs18}) also increases.

Irrespective of the value of $\Omega$, when (\ref{bs18}) is
applied to the combined strain rate field $S_{ij}(\bf{x})$ in
(\ref{burg3}) and (\ref{burg45}), if
$\tilde{R}\equiv(R/\lambda_{\nu}) \rightarrow \infty$ and all
orders $n$ are retained then the resulting $S^{B}_{ij}(\bf{x})$
should recover the background strain field, namely
\begin{equation}\label{convergence}
S_{r\theta}^B \rightarrow 0
\end{equation}
for all $\bf{x}$, and the vorticity should show alignment with the most extensional
principal axis of $S_{ij}^{B}$.
For finite $R/\lambda_{\nu}$ and various orders $n$, the convergence of
$S_{ij}^{B}$ from (\ref{bs18}) to this background strain field is examined below.

\begin{figure}[b]
\includegraphics[width=3.5in]{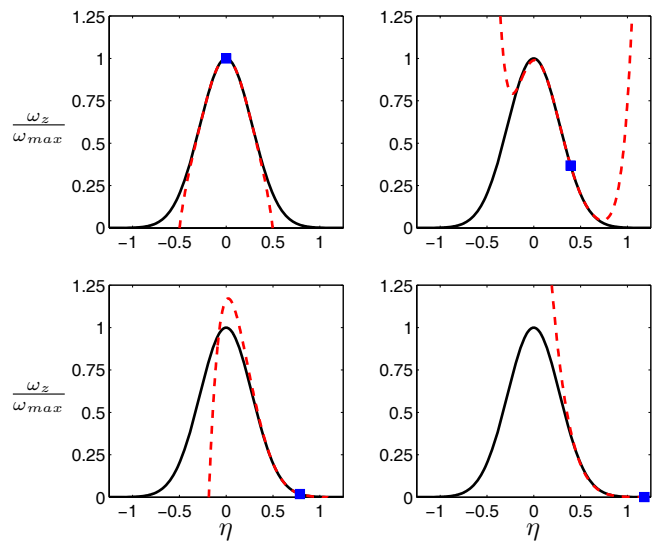}
\caption{Accuracy of the Taylor expansion for the local vorticity in (\ref{bs8}) for a Burgers vortex,
showing results for 6th order approximation. In each panel, solid black curve shows actual vorticity
profile, and red dashed curve gives approximated vorticity from derivatives at location marked by
square.}
\label{TaylorFit}
\end{figure}
\begin{figure}[t]
\includegraphics[width=3in]{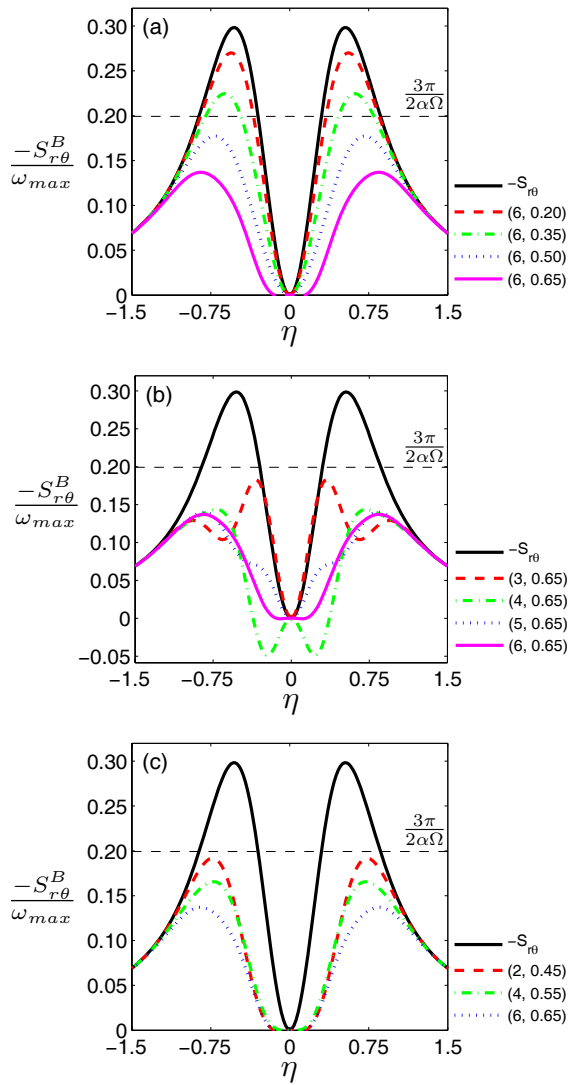}
\caption{\label{fig:wide}Convergence of background strain rate
field $S_{ij}^{B}(\bf{x})$ for a Burgers vortex, obtained from
total strain rate field $S_{ij}(\bf{x})$ using (\ref{bs18}) for
various $(n,\tilde{R})$ combinations, where $\tilde{R}\equiv
R/\lambda_\nu$. Shown are effects of increasing $\tilde{R}$ for
fixed $n = 6$ ($top$), increasing $n$ for fixed $\tilde{R} = 0.65$
($middle$), and increasing $n$ and $\tilde{R}$ simultaneously
($bottom$). The dashed horizontal lines follow from
(\ref{switch1}) and (\ref{axisswitchparam}).} \label{burgf3}
\end{figure}

\subsection{Convergence of the Background Strain}
The accuracy with which (\ref{bs18}) can recover the background
strain field $S_{ij}^B(\textbf{x})$ that acts on a concentrated
vortical structure depends on how well the expansion in
(\ref{bs8}) represents the vorticity field within the local
spherical neighborhood $R$.  Figure \ref{TaylorFit} shows the
results of a local sixth-order Taylor series approximation for the
vorticity in (\ref{burg1}) at various radial locations across the
Burgers vortex.  In each panel, the blue square marks the location
$\bf{x}$ at which the sphere is centered, and the red dashed curve
shows the resulting Taylor series approximation for the vorticity.
On the axis of the vortex, the approximated vorticity field
correctly accounts for most of the circulation in the vortex, and
thus the induced strain field from the vortex will be reasonably
approximated. Off the axis, the approximation becomes increasingly
poorer, but the $1/r^2$ decrease in the Biot-Savart kernel in
(\ref{bs1a}) nevertheless renders it adequate to account for most
of the vortex-induced strain rate field.  At the largest radial
location, corresponding to the bottom right panel of Fig.\
\ref{TaylorFit}, the approximation becomes relatively poor,
however at large $\eta$ values the vortex-induced strain is
sufficiently small that it is unlikely to lead to alignment
switching for typical $\Omega$ values.

Figure \ref{burgf3} shows the shear component
$S_{r\theta}^{B}(\eta)$ of the background strain rate tensor
obtained via (\ref{bs18}) for various $n$ and $\tilde{R}$ as a
function of $\eta$. In each panel, the black curve shows the total
strain rate $S_{ij}(\eta)$ and the colored curves show the
background strain rate $S_{ij}^{B}(\eta)$ from (\ref{bs18}) for
the $(n,\tilde{R})$ combinations listed. The horizontal dashed
line corresponding to $\Omega = (3/2)\Omega^{*}$ reflects the
relative vortex strength, and shows the range of $\eta$ where the
anomalous alignment switching occurs due to the vortex-induced
strain field.  Wherever the $-S^B_{r\theta}$ curves are above this
line, the vorticity there will be aligned with the intermediate
principal axis of the $combined$ strain rate tensor $S_{ij}$.
Figure \ref{burgf3}$(a)$ examines the effect of increasing the
radius $\tilde{R}$ of the spherical region for fixed order $n =
6$. It is apparent that with increasing $\tilde{R}$ the resulting
$-S_{r\theta}^{B}$ converges toward the correct background strain
field in (\ref{convergence}). For the value of $\Omega$ shown, it
can be seen that for $R \gtrsim 0.5 \lambda_{\nu}$ the resulting
$S_{r\theta}^{B}$ is everywhere below the horizontal dashed line,
indicating that the vorticity everywhere is aligned with the most
extensional principal axis of the resulting background strain rate
tensor $S_{ij}^{B}(\textbf{x})$ from (\ref{bs18}).

In Fig.\ \ref{burgf3}$(b)$ similar results are shown for the
effect of increasing the order $n$ of the expansion for the
vorticity field for fixed $\tilde{R} = 0.65$. It is apparent that
the effect of $n$ is somewhat smaller than for $\tilde{R}$ in
Fig.\ \ref{burgf3}$(a)$.  Moreover, the results suggest that the
series in (\ref{bs18}) alternates with increasing order $n$.  For
this $\Omega$ and  $\tilde{R}$, even $n = 3$ is seen to be
sufficient to remove most of the vortex-induced shear strain, and
thus reduce $S_{r\theta}^{B}(\textbf{x})$ below the horizontal
dashed line. For these parameters, the $S_{r\theta}^{B}$ field
from (\ref{bs18}) would thus reveal alignment of the vorticity
with the most extensional principal axis of the background strain
tensor throughout the entire field.

Figure \ref{burgf3}$(c)$ shows the combined effects of increasing
both $\tilde{R}$ and $n$, in accordance with the expectation that
larger $\tilde{R}$ should require a higher order $n$ to adequately
represent the vorticity field within the spherical region. The
shear strain rate field shows convergence to the correct
background strain field in (\ref{convergence}). The convergence of
the shear strain rate $S_{r\theta}^B(\textbf{x})$ to zero in the
vicinity of the vortex core is of particular importance. The
systematic reduction in the peak remaining shear stress indicates
that, even for increasingly stronger vortices or increasingly
weaker background strain fields as measured by $\Omega$, the
resulting $S_{r\theta}^B(\textbf{x})$ from (\ref{bs18}) will
reveal the alignment of all the vorticity in such a structure with
the most extensional principal strain axis of the background
strain field.

\section{Vorticity Alignment in Turbulent Flows}
Having seen in the previous Section how (\ref{bs18}) is able to
reveal the expected alignment of vortical structures with the most
extensional eigenvector of the {\it background} strain rate in
which they reside, in this Section we apply it to obtain insights
into the vorticity  alignment in turbulent flows. In particular,
we examine the alignment at every point ${\bf x}$ of the vorticity
$\bm{\omega}$ relative to the eigenvectors of the total strain
rate tensor field $S_{ij}(\bf{x})$ and those of the background
strain field data $S_{ij}^{B}(\bf{x})$. This analysis uses data
from a highly-resolved, three-dimensional, direct numerical
simulation (DNS) of statistically stationary, forced, homogeneous,
isotropic turbulence \cite{Schumacher2007, Schumacher2007a}. The
simulations correspond to a periodic cube with sides of length of
$2\pi$ resolved by $2048^3$ grid points.  The Taylor microscale
Reynolds number $R_{\lambda}$ is 107.

The DNS data were generated by a pseudospectral method with a
spectral resolution that exceeds the standard value by a factor of
eight.  As a result, the highest wavenumber corresponds to
$k_{max}\eta_K=10$, and the Kolmogorov lengthscale
$\eta_{K}=\nu^{3/4}/\langle\epsilon\rangle^{1/4}$ is resolved with
three grid spacings.  This superfine resolution makes it possible
to apply the result in (\ref{bs18}) for relatively high orders
$n$, which require accurate evaluation of high-order derivatives
of the DNS data. In Schumacher {\it et al.} \cite{Schumacher2007}
it was demonstrated that derivatives up to order six are
statistically converged.  More details on the numerical
simulations are given in Refs.
\cite{Schumacher2007,Schumacher2007a}.

\begin{figure}[t]
\includegraphics[width=3.5in]{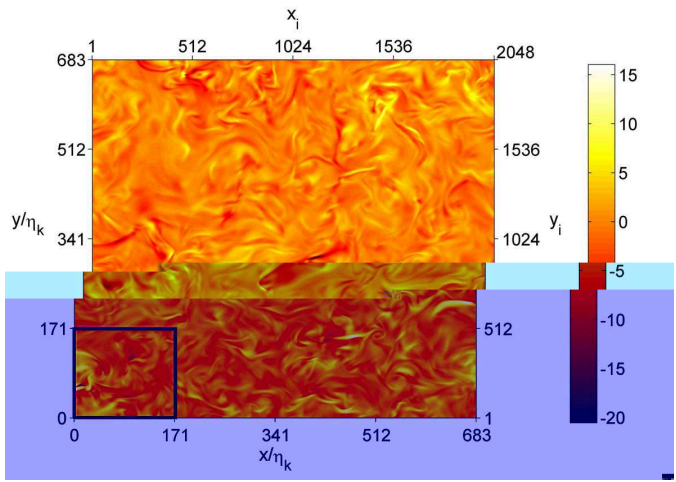}
\caption{Instantaneous snapshot of total strain rate component
field $S_{12}(\bf{x})$ in a two-dimensional slice through a
highly-resolved three-dimensional $(2048^3)$ DNS of homogeneous,
isotropic turbulence \cite{Schumacher2007,Schumacher2007a}. Axes
are given both in grid coordinates ($i=1\ldots2048$) and
normalized by the Kolmogorov length $\eta_K$. Box indicates region
in which background strain rate field $S_{ij}^{B}(\bf{x})$ is
computed in Fig. \ref{burgf5}.} \label{burgf4}
\end{figure}

Figure \ref{burgf4} gives a representative sample of the DNS data, where the instantaneous shear
component $S_{12}$ of the total strain rate tensor field $S_{ij}(\bf{x})$ is shown in a typical
two-dimensional intersection through the $2048^3$ cube.  The data can be seen to span nearly 700 $\eta_K$ in each direction.  The $512^2$ box at the lower left of Fig.\ \ref{burgf4} is used here to obtain initial results for alignment of the vorticity with the eigenvectors of the background strain rate tensor.

\begin{figure}[b]
\includegraphics[width=3.5in]{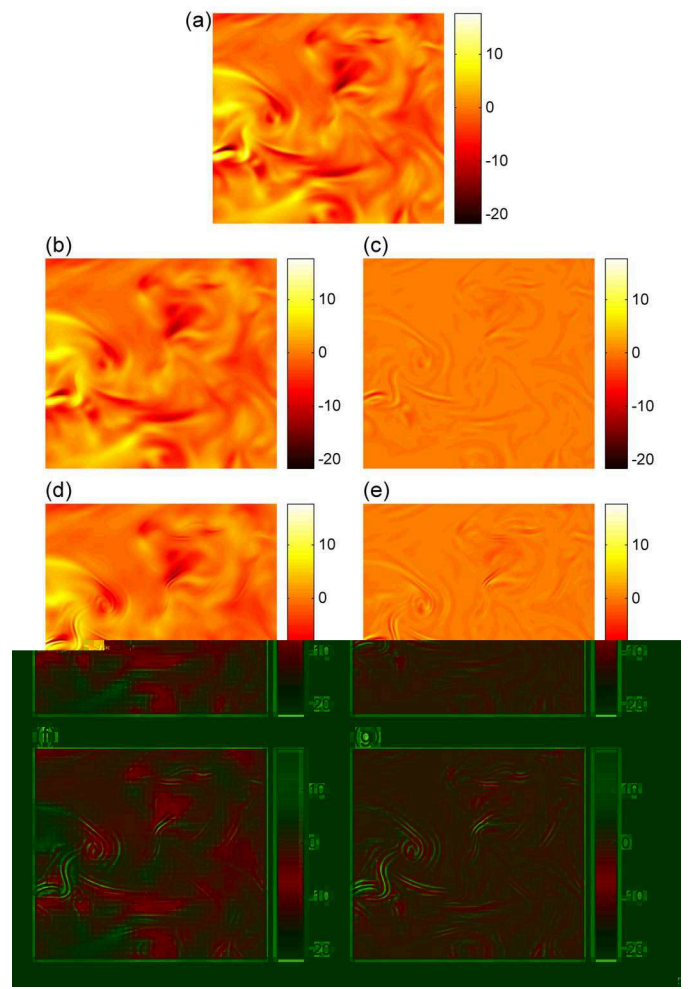}
\caption{Total strain rate component field $S_{12}(\bf{x})$ ($a$),
with corresponding results from (\ref{bs18}) for nonlocal
(background) field $S_{12}^{B}(\bf{x})$ ($left$) and local field
$S_{12}^{R}(\bf{x})$ ($right$) for $\left(R/\eta_K\right)=2.5$
($b,c$), $\left(R/\eta_K\right)=3.5$ ($d,e$), and
$\left(R/\eta_K\right)=4.5$ ($f,g$), all with $n=3$.}
\label{burgf5}
\end{figure}

The background strain rate tensor field $S_{ij}^{B}(\bf{x})$ is
first extracted via (\ref{bs18}) from $S_{ij}({\bf x})$ for $n=3$
and various $\left(R/\eta_K\right)$. Higher-order evaluation of
the background strain rate is not feasible, as the results in
Ref.\ \cite{Schumacher2007} show that only spatial derivatives of
the velocity field up to order six can be accurately obtained from
these high-resolution DNS data. For $n=4$, the expansion in
(\ref{bs18}) involves seventh-order derivatives of the velocity
field, and the background strain evaluation becomes limited due to
the grid resolution. The results are shown and compared in Fig.\
\ref{burgf5}, where the shear component $S_{12}$ of the full
strain rate tensor is shown at the top, and the corresponding
nonlocal (background) component $S_{12}^{B}$ and local component
$S_{12}^{R}$ are shown, respectively, in the left and right
columns for $\left(R/\eta_K\right)=2.5$ ($top$ $row$), $3.5$
($middle$ $row$), and $4.5$ ($bottom$ $row$). Consistent with the
results from the Burgers vortex in Fig.\ \ref{burgf3}, as
$\left(R/\eta_K\right)$ increases the magnitude of the extracted
local strain rate in the right column increases. However, for the
largest $\left(R/\eta_K\right)=4.5$ case, $n=3$ appears to be too
small to adequately represent the local vorticity field. This
leads to truncation errors which are manifested as strong ripples
in the background and local strain fields (see panels (\textit{f})
and (\textit{g})).


\begin{figure*}[t]
\includegraphics[width=5in]{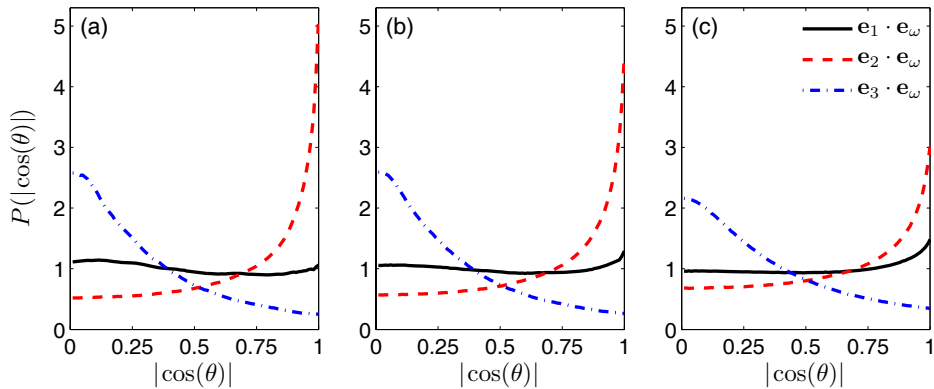}
\caption{Probability densities of alignment cosines for the
vorticity with the eigenvectors of the strain rate tensor, showing
results for $S_{ij}$ ($a$) and for $S_{ij}^B$ using
$\left(R/\eta_K\right)=2.5$ with $n=3$ ($b$) and
$\left(R/\eta_K\right)=3.5$ with $n=3$ ($c$).} \label{alignments}
\end{figure*}

The results in Fig.\ \ref{burgf5} thus indicate that radii up to
$\left(R/\eta_K\right)=3.5$ in combination with $n=3$ can be used
to assess alignment of the vorticity vector with the eigenvectors
of the background strain rate field. Figure \ref{alignments} shows
the probability densities of the alignment cosines for the
vorticity vector with the total strain rate tensor and with the
background strain rate tensors from (\ref{bs18}). We compare
$S_{ij}$ (Fig.\ \ref{alignments}\textit{a}) with $S_{ij}^B$ for
$\left(R/\eta_K\right)=2.5, n=3$ (Fig.\
\ref{alignments}\textit{b}) and $S_{ij}^B$ for
$\left(R/\eta_K\right)=3.5, n=3$ (Fig.\
\ref{alignments}\textit{c}). The results for alignment with the
total strain rate tensor are essentially identical to the
anomalous alignment seen in numerous other DNS studies
\cite{Ashurst1987,She1991,Nomura1998} and experimental studies
\cite{Tsinober1992,Buch1996,Su1996,Zeff2003,Mullin2006}, which
show the vorticity to be predominantly aligned with the
eigenvector corresponding to the intermediate principal strain
rate.  However, the results for the Burgers vortex in the previous
section show that such anomalous alignment with the eigenvectors
of the total strain rate tensor is expected when the local vortex
strength parameter $\Omega$ is sufficiently large to cause
alignment switching.

By comparison, the results in Fig.\ \ref{alignments} (\textit{b})
and (\textit{c}) obtained for the alignment cosines of the
vorticity vector with the $background$ strain rate tensor
$S_{ij}^{B}$ from (\ref{bs18}) show a significant decrease in
alignment with the intermediate eigenvector, and an increase in
alignment with the most extensional eigenvector.  While data in
panel (\textit{b}) show only a slight change compared to those in
(\textit{a}), the results in panel (\textit{c}) demonstrate that
our decomposition can indeed diminish the anomalous alignment
significantly. This is consistent with the results for the Burgers
vortex in the previous Section, and with the hypothesis that the
alignment switching mechanism due to the local contribution
$S_{ij}^{R}$ to the total strain rate tensor is the primary reason
for the anomalous alignment seen in earlier studies.  It is also
consistent with the expected equilibrium alignment from
(\ref{bs0}).  While a more detailed study is needed to examine
possible nonequilibrium contributions to the alignment
distributions associated with eigenvector rotations of the
background strain field, as well as to definitively determine the
$R$ and $n$ convergence of the background strain rate tensor in
Fig.\ \ref{burgf5}, the present findings support both the validity
of the result in (\ref{bs18}) for extracting the background strain
rate tensor field $S_{ij}^{B}(\bf{x})$ from the total strain rate
tensor field $S_{ij}(\bf{x})$, and the hypothesis that at least
much of the anomalous alignment of vorticity in turbulent flows is
due to the differences between the total and background strain
rate tensors and the resulting alignment switching noted herein.

\section{Concluding Remarks}
We have developed a systematic and exact result in (\ref{bs18})
that allows the local and nonlocal (background) contributions to
the total strain rate tensor $S_{ij}$ at any point $\bf{x}$ in a
flow to be disentangled.  The approach is based on a series
expansion of the vorticity field in a local spherical neighborhood
of radius $R$ centered at the point $\bf{x}$.  This allows the
background strain rate tensor field $S_{ij}^{B}(\bf{x})$ to be
determined via a series of increasingly higher-order Laplacians
applied to the total strain rate tensor field $S_{ij}(\bf{x})$.
For the Burgers vortex, with increasing radius $R$ relative to the
local gradient lengthscale $\lambda_{\nu}$ and with increasing
order $n$, we demonstrated convergence of the resulting background
strain tensor field to its theoretical form.  We also showed that
even with limited $R$ and $n$ values, the local contribution to
the total strain rate tensor field can be sufficiently removed to
eliminate the anomalous alignment switching throughout the flow
field. This conclusion is expected to also apply to the more
realistic case of a non-uniformly stretched vortex where
$S_{zz}=f(z)$
\cite{Gibbon1999,Ohkitani2002,Cuypers2003,Rossi2004}.

Consistent with the results for the Burgers vortex, when
(\ref{bs18}) was used to determine the background strain rate
tensor field $S_{ij}^{B}(\bf{x})$ in highly-resolved DNS data for
a turbulent flow, the anomalous alignment seen in previous DNS and
experimental studies was substantially reduced.  We conclude that
(\ref{bs18}) allows the local background strain rate tensor to be
determined in any flow. Furthermore, we postulate that the
vorticity vector field in turbulent flows will show a
substantially preferred alignment with the most extensional
principal axis of the background strain rate field, and that at
least much of the anomalous alignment found in previous studies is
simply a reflection of the alignment switching mechanism analyzed
in Section III and conjectured by numerous previous investigators.

Lastly, the result in (\ref{bs18}) is based on a Taylor series
expansion of the vorticity within a spherical neighborhood of
radius $R$ around any point $\bf{x}$.  Such an expansion
inherently involves derivatives of the total strain rate tensor
field, which can lead to potential numerical limitations.  If
larger $R$ and correspondingly higher orders $n$ are needed to
obtain accurate evaluations of background strain rate fields, then
otherwise identical approaches based on alternative expansions may
be numerically advantageous.  For instance, an expansion in terms
of orthonormal basis functions allows the coefficients to be
expressed as integrals over the vorticity field within $r \le R$,
rather than as derivatives evaluated at the center point $\bf{x}$.
(For example, wavelets have been used to test alignment between
the strain rate eigenvectors and the vorticity gradient in
two-dimensional turbulence \cite{Protas2002}.)  This would allow a
result analogous to (\ref{bs18}) that can be carried to higher
orders with less sensitivity to discretization error.  The key
conclusion, however, of the present study is that it is possible
to evaluate the background strain tensor following the general
procedure developed herein, and that when such methods are applied
to assess the background strain rate fields in turbulent flows
they reveal a substantial increase in the expected alignment of
the vorticity vector with the most extensional principal axis of
the background strain rate field.

\begin{acknowledgments}
PH and WD acknowledge support from the Air Force Research
Laboratory (AFRL) under the Michigan AFRL Collaborative Center for
Aeronautical Sciences (MACCAS), and by the National Aeronautics \&
Space Administration (NASA) Marshall and Glenn Research Centers
and the Department of Defense (DoD) under the NASA Constellation
University Institutes Project (CUIP) under Grant No. NCC3-989. JS
acknowledges support by the German Academic Exchange Service
(DAAD) and by the Deutsche Forschungsgemeinschaft (DFG) under
grant SCHU 1410/2. The direct numerical simulations have been
carried out within the Deep Computing Initiative of the DEISA
consortium on 512 CPUs of the IBM-p690 cluster JUMP at the John
von Neumann Institute for Computing at the Research Centre
J\"ulich (Germany).

\end{acknowledgments}

\end{document}